\documentclass[twocolumn,showpacs,showkeys,superscriptaddress,aps,pra]{revtex4-2}
\usepackage{amsfonts}
\usepackage{amssymb}
\usepackage{amsmath}
\usepackage{xcolor}
\usepackage{graphicx}
\usepackage{dcolumn}
\usepackage{bm}
\usepackage{subfigure}
\usepackage{ulem} 
\begin{document}

\title{Extreme events in a broad-area semiconductor laser with coherent injection}

\author{Cristina Rimoldi}
 \affiliation{Dipartimento di Elettronica e Telecomunicazioni, Politecnico di Torino, Torino, Italy}
\author{Mansour Eslami}
 \affiliation{Department of Physics, University of Guilan, P.O. Box 41335-1914 Rasht, Iran}
\author{Franco Prati}
\affiliation{Dipartimento di Scienza e Alta Tecnologia, Universit\`{a} dell’Insubria, Via Valleggio 11, 22100 Como, Italy}
\author{Giovanna Tissoni}
 \email{giovanna.tissoni@inphyni.cnrs.fr}
 \affiliation{Universit\'{e} C\^{o}te d’Azur, CNRS, Institut de Physique de Nice, Valbonne, France}

\date{\today}
\begin{abstract}

Spatiotemporal extreme events are interesting phenomena, both from a fundamental point of view, as manifestations of complexity in dynamical systems, and for their possible applications in different research fields.
Here, we present some recent results about extreme events in spatially extended semiconductor laser systems (broad-area VCSELs) with coherent injection. 
We study the statistics of spatiotemporal intensity peaks occurring in the transverse (x,y) section of the field perpendicular to the light propagation direction and identify regions in the parameter space where extreme events are more likely to occur. 
Searching for precursors of these phenomena, we concentrate, on one hand, on the spatiotemporal dynamics of the field phase and in particular on the presence of optical vortices in the vicinity of an extreme event. On the other hand, we focus on the laser gain dynamics and the phase space trajectories of the system close to the occurrence of an extreme event. Both these complementary approaches are successful and allow us to shed some light on potential prediction strategies.

\end{abstract}

\pacs{Valid PACS appear here}
\maketitle

\section{Introduction}

In the last decades, extreme events have been under study in many different physical systems, from hydrodynamics to optics. In this last field, they have been defined as optical rogue waves (RWs) \cite{Solli2007} and were the topic of a vast amount of literature in many different optical systems (for a review, see Refs. \cite{Onorato2013,Dudley2014,Akhmediev_2016} and references therein). Optical fibers and fiber lasers have been identified since the beginning as systems of choice for optical rogue wave studies due to their natural longitudinal extension (see, for instance, Refs. \cite{Kibler2010,ErkintaloPRL2011,Kibler2012} for conservative cases and Refs. \cite{Wabnitz2017,Coulibaly2019} for dissipative rogue waves in resonators and laser devices).\\ 
Semiconductor systems have also emerged as experimentally convenient test-beds for the analysis of extreme phenomena. For instance, low-dimensional semiconductor systems, in which the wave envelope is severely constrained by boundary conditions, served to demonstrate that the emergence of rogue events can be associated with an external crisis in a chaotic regime \cite{Bonatto2011}, thus showing the deterministic character of these extreme events. This deterministic nature was also revealed in the analysis of delayed feedback semiconductor lasers where the effect of increased noise reduces the probability of rogue waves by preventing the dynamics from approaching the narrow path in the phase space that leads to extreme pulses \cite{Zamora-Munt2013}.\\
Various mechanisms have been identified for the formation of rogue waves, from modulational instability \cite{Dudley2009} to soliton and breather occurrence \cite{Kibler2010} and competition of extended structures associated to opposite signs of nonlinearity \cite{EslamiOppoRWPRA,EslamiOppoEITPRA}. Rogues waves have also been reported from the interaction of optical vortices with opposite chiral charges \cite{Gibson2016}. However, a unifying mechanism for extreme event generation in the optical context remains undiscovered.\\ 
Very recently, extreme events were studied both experimentally and numerically \cite{Selmi2016,Coulibaly2017} in the intensity of the electric field emitted by a monolithic broad-area vertical cavity surface-emitting laser (VCSEL) with a saturable absorber in case of a linear pump (which reduces the number of transverse dimensions to one) and spatiotemporal
chaos was claimed to be at the dynamical origin of these objects. The very same system has been under investigation in a 2D configuration, where spatiotemporal extreme events have been identified as maxima of the field intensity in the three-dimensional space (x,y,t) \cite{Rimoldi2017} and can be controlled through harmonic pump modulation \cite{Talouneh2020}. Extreme events have been analyzed following their various definitions and through different RW indicators, and the best parameter choice to observe them has been identified. Furthermore, it has been possible to determine the typical temporal and spatial size (FWHM) expected for such extreme events and to compare it with stationary and oscillating solitons. As suggested in Ref. \cite{Bonazzola2013} for a similar system, we believe that two-dimensional spatial effects play a crucial role in the formation of extreme events.\\
Spatially extended semiconductor lasers in a macroscopic ring cavity configuration and with coherent injection have also been extensively studied in very recent years \cite{Gustave2017,Rimoldi2017a} and extreme events have been identified in different parametric regimes. Interestingly, the dynamics of the electric field phase seems to play an important role in the generation of extreme events, allowing to draw a natural analogy with the particular kind of dissipative solitons existing in these systems, i.e. phase solitons \cite{Gustave2015}, with which extreme events share their chiral nature. \\
In this paper we consider a broad-area semiconductor laser (VCSEL) with coherent injection \cite{Hachair2006} and focus on  the (spatiotemporal) `turbulent' regime occurring for low values of injection where the lower branch of the homogeneous stationary state (HSS) is Hopf unstable. A similar regime has been studied in \cite{Gibson2016} for a generic class-A laser when the cavity and the atomic medium are resonant, i.e. with $\Delta=\theta=0$, and characterized as vortex turbulence. We would like to stress that, even if the model here preserves some of the features illustrated in \cite{Gibson2016} (for instance we will see that the phase of the electric field still seems to play an important role), the system dynamics is dramatically different. This is because (i) the insertion of the linewidth enhancement factor $\alpha$ for semiconductor lasers in the model equations, playing the role of the atomic detuning $\Delta$, changes the degree of complexity of the observed turbulence and (ii) the carrier density dynamics is not fast enough to be (adiabatically or non-adiabatically) eliminated, which leads to interesting consequences in the prediction of extreme events \cite{Alvarez2017}.\\
Finally, some dynamical similarities can be drawn with the model studied in \cite{Gustave2017, Rimoldi2017a}, for a semiconductor ring laser with injection. As mentioned already in \cite{Anbardan2018}, for a model fairly similar to the one utilized here, the exponential decay of the merging time of two cavity solitons (CSs) as a function of their initial distance somehow resembles the transition time necessary for two phase solitons carrying a single chiral charge along the propagation direction to merge in a phase soliton complex carrying a double charge. The main difference between these two cases consists in the fact that in \cite{Gustave2017, Rimoldi2017a} the system (single) spatial dimension was given by the propagation along the ring cavity while here the two spatial dimensions describe the laser transverse plane and are perpendicular to the propagation direction.\\
The paper is structured as follows. In Section II, we illustrate the theoretical model and show the main results of a linear stability analysis of the homogeneous steady state solution. In Section III, we show the results of numerical simulations and statistically prove the presence of extreme events in the system and study the mechanism leading to their generation, with a particular focus on optical vortices. In Section IV, we study rogue wave occurrence in terms of optical gain dynamics and in Section V, we prove that these events persist in presence of a finite pump profile. Finally in Section VI and VII, we discuss our work, draw our conclusions, and suggest some possible extensions of these results.

\section{Theory}
The model describing a broad-area semiconductor laser (VCSEL) with injection is given by the following set of rate equations
\begin{eqnarray}\label{eq:model}
\dot{E}\hspace{-0.7mm}&=&\hspace{-0.7mm}\sigma\hspace{-0.7mm}\left[E_I-(1+i\theta)E+(1-i\alpha)DE+(d+i)\nabla_\bot^2E\right]\nonumber\\
\dot{D}\hspace{-0.7mm}&=&\hspace{-0.7mm}\mu-D\left(1+\left|E\right|^2\right)\,,
\end{eqnarray}
which derives from a set of effective Maxwell-Bloch equations known for properly describing this kind of system \cite{Hachair2006}, where the polarization of the semiconductor medium has been adiabatically eliminated. The dynamical variables $E$ and $D$ are, respectively, the slowly varying electric field and the semiconductor carrier density. $E_I$ is the injected field amplitude, $\theta$ represents the detuning between the cavity and the injected field frequencies, $\alpha$ is the linewidth enhancement factor typical of semiconductor lasers, $\sigma$ is the ratio of carrier lifetime to photon lifetime and $\mu$ is the injection current (the free running laser threshold being $\mu_{thr}=1$). For $\theta+\alpha=0$ the injected field is resonant with the free-running laser frequency. Finally, $d$ is a diffusive term phenomenologically introduced to take into account a finite linewidth for the laser gain. A detailed analytic derivation for this term from a nonstandard adiabatic elimination of the polarization can be found in \cite{Fedorov2000} for a broad-area laser with a saturable absorber. Time is scaled to the carrier lifetime ($\approx 1$ ns) and space is scaled to the square root of the diffraction parameter (leading to a space unit of about 4 $\mu$m).\\
\begin{figure}[t!]
\centerline{\subfigure[]
{\includegraphics[width=.55\columnwidth]{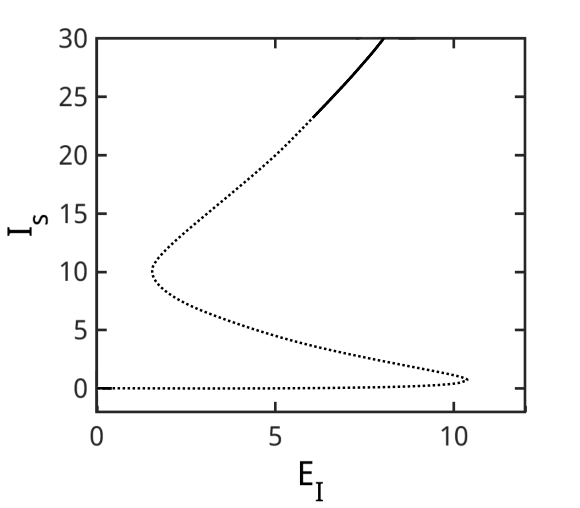}}
\subfigure[]
{\includegraphics[width=.5\columnwidth]{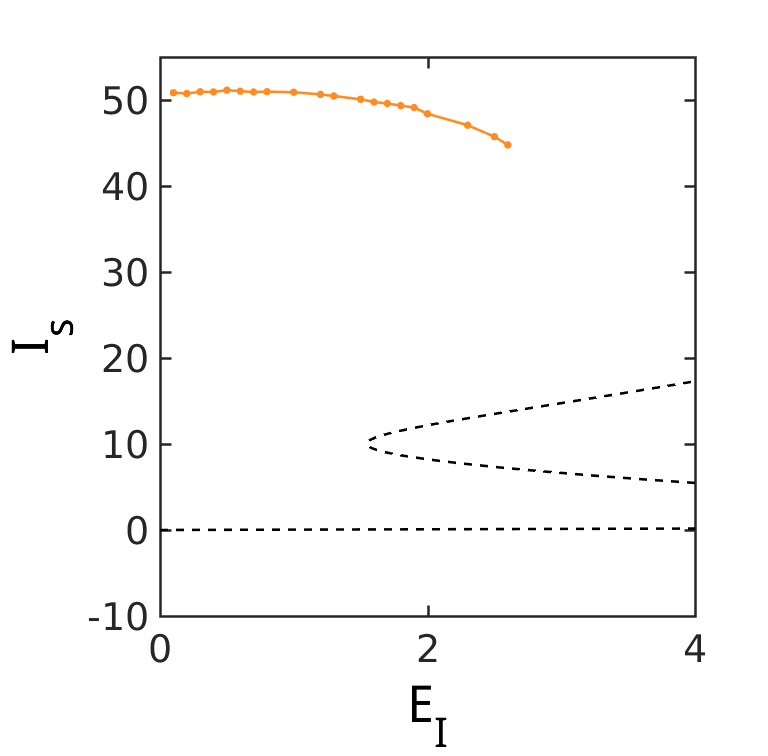}}}
\centerline{\subfigure[]
{\includegraphics[width=.5\columnwidth]{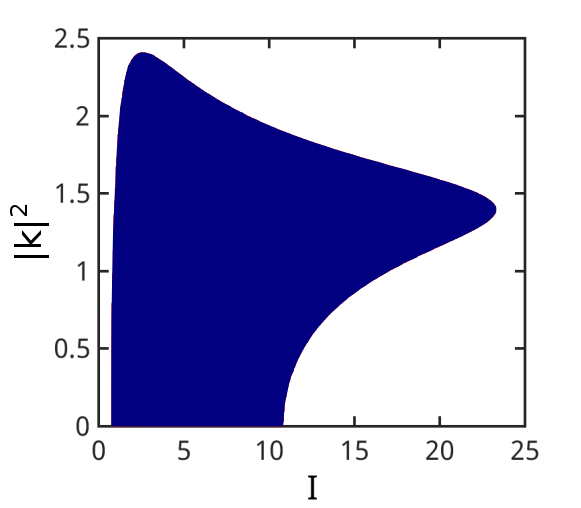}}
\subfigure[]
{\includegraphics[width=.5\columnwidth]{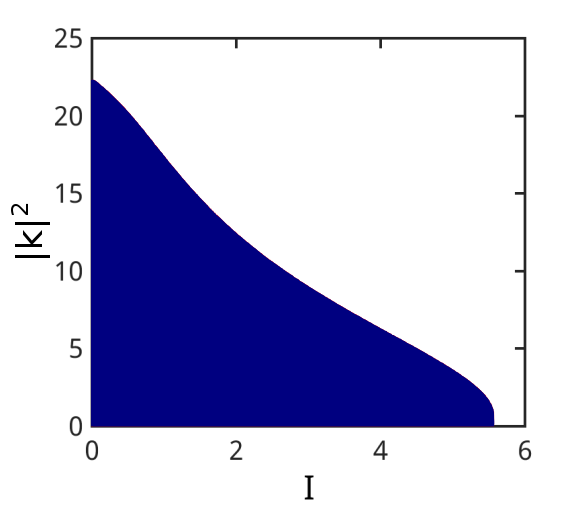}}}
\caption{(a) Homogeneous stationary solution (HSS) for the laser intensity $I_s=|E_s|^2$ as a function of $E_I$. (b) The same HSS plotted together with the branch for the stable `turbulent' solution (orange line). Stationary (c) and Hopf (d) instability domains for the fixed parameters $\alpha=4$, $\theta=-2$, $\mu=6$, $d=0.01$ and $\sigma=400$. 
}
\label{fig:HSS}
\end{figure}
The homogeneous stationary solution (HSS) for the model in Eq. \eqref{eq:model} reads $E_s=|E_s|\exp(i\phi_s)$ and $D_s=\mu/(1+|E_s|^2)$, with
\begin{eqnarray}
E_I^2&=&|E_s|^2\left[\left(1-D_s\right)^2+\left(\theta+\alpha D_s\right)^2\right]\label{eq:HSS}\\
\phi&=&\arctan\left(\frac{\theta+\alpha D_s}{D_s-1}\right)\,.
\end{eqnarray}
In Fig. \ref{fig:HSS}(a) we plotted Eq. \eqref{eq:HSS} for the laser intensity as a function of the injection amplitude for the following choice of parameters: $\alpha=4$, $\theta=-2$, $\mu=6$ and $\sigma=400$. In particular, the given value of $\sigma$ implies the values $\tau_p=2.5$ ps for the photon lifetime and $\tau_c=1$ ns for the carrier recombination time, according to literature and experimental measurements \cite{Hachair2006}. The diffusion term $d=0.01$ has been chosen as the smallest possible value to avoid self-collapsing. In Fig. \ref{fig:HSS}(c,d) we illustrated, respectively, the stationary and Hopf instability domains in the plane defined by the square modulus of the transverse wave vector $|k|^2=k_x^2+k_y^2$ and the laser intensity. According to Fig. \ref{fig:HSS}(c) and (d) the HSS is stationary unstable in the negative slope branch (plane wave instability, $|k| = 0$) and in part of the upper branch (modulational or Turing instability, with  $|k| \neq 0$), and Hopf unstable in the lower branch.\\ 
For a slightly different model and a lower current value but still above the free running laser threshold, the interaction of cavity solitons has been recently investigated in presence of a lower HSS branch that was only partially Hopf unstable and compared with similar results for soliton merging time obtained in hydrophobic materials \cite{Anbardan2018}. Here we focus instead on the model dynamics when the system undergoes Hopf instability in its entire lower branch (i.e. for low values of injection). We would also like to point out that, at difference from \cite{Rimoldi2017} where parameter regions of coexistence between cavity solitons (CSs) and spatiotemporal chaos (STC) could be found, in the present context CSs exist only where (part of) the lower HSS branch is stable, as in \cite{Anbardan2018}, given the requirement for a stable background.\\  
Although stable CSs have been reported to exist atop temporally (due to Hopf instability) \cite{Hachair2006, Eslamiphysica13} and/or spatially (due to Turing instability) \cite{EslamiOppo19} unstable backgrounds, in the present model a profile comparison with CSs is not possible since extreme events are found for values of current for which coexistence of a higher-intensity spatially-modulated branch with a homogeneous low-intensity branch, essential to creation of CSs, is absent.
For the choice of parameter values indicated above, the system exhibits a turbulent behavior in the branch highlighted in orange in Fig. \ref{fig:HSS}(b) for $0 < E_I< 2.6$ where we illustrated the temporal average of the spatial maxima recorded in the transverse plane during 25 ns-long simulations. In the figure, a decrease in the value of averaged maxima for higher values of injection is evident.\\

\section{Extreme events and their statistics}
In order to characterize the presence of extreme events in the system, we run simulations for the set of parameters mentioned in the previous section, fixed $\mu=6$ and varied the injection amplitude $E_I$ (and vice versa, fixed $E_I=0.5$ and varied $\mu$). For each simulation we initialized the system long enough to overcome any transient behavior and then performed data acquisition for a 25 ns-long window (unless stated otherwise). The sampling rate in the transverse plane was 1 ps on a grid size of 256$\times$256 pixels (corresponding approximately to 256$\times$256 $\mu$m$^2$ when a space step of 0.25 is adopted, as in most of our simulations).\\

\begin{figure}[t!]
\centerline{\subfigure[]
{\includegraphics[width=0.4\textwidth]{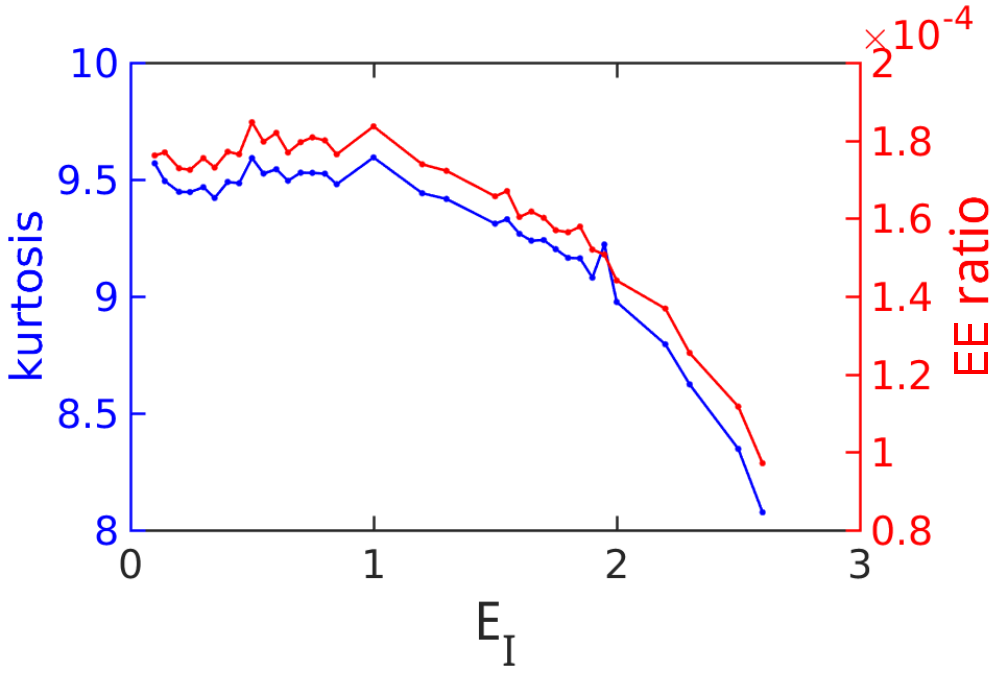}}}
\centerline{\subfigure[]
{\includegraphics[width=0.8\columnwidth]{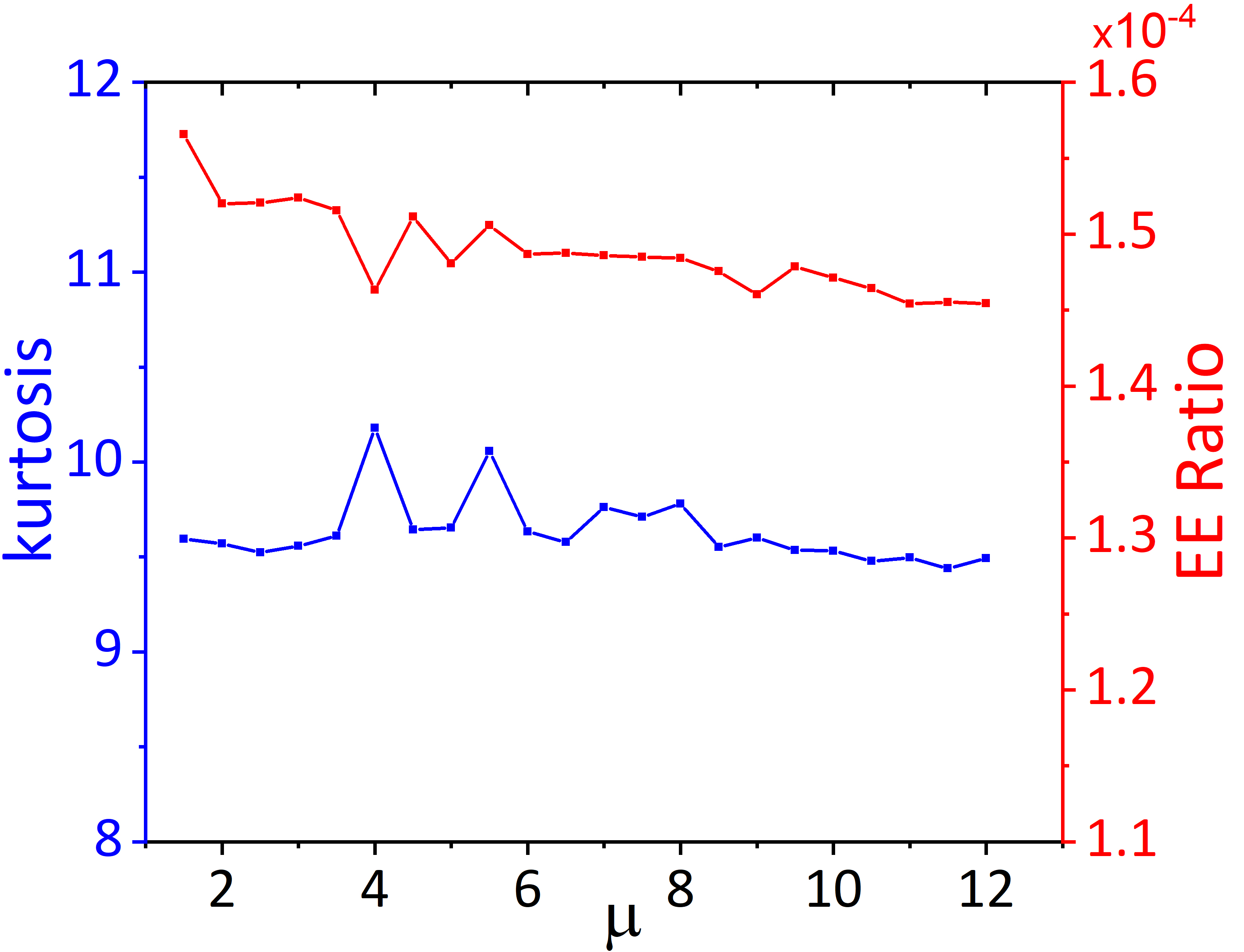}}}
\caption{Kurtosis ($\mathcal{K}$), in blue, and extreme event (EE) ratio, in red, of the total intensity PDF for simulations with different (a) injection amplitudes $E_I$ and (b) pump currents, respectively for $\mu=6$ and $E_I=0.5$. From these two indicators we can observe that extreme events are more likely to be observed for low values of $E_I$, while variations of pump current for fixed injection do not play a significant effect on extreme event occurrence.}
\label{fig:kurt_RW_ratio}
\end{figure}

An estimation of the data extremeness can be obtained by computing the probability density function (PDF) of all the values explored by the electric field intensity $I$ during the simulations. As discussed in literature \cite{Onorato2013}, if the {\it `surface elevation'} of the real and imaginary parts of the electric field follows a Gaussian statistics, this is reflected in a negative exponential behavior as $\exp(-I_{tot}/\langle I_{tot}\rangle)/\langle I_{tot}\rangle$ for the PDF of the total intensity. Hence, any positive deviation from a negative exponential statistics has to be considered a signature of the presence of extreme events in the system. An indicator often used to characterize the extreme nature of the turbulent regime as a function of the injection is the kurtosis ($\mathcal{K}$), that is the ratio of the fourth momentum about the mean of the data to the square of its variance. This statistical tool gives a measure of the tails of a distribution. Even though the kurtosis is not to be interpreted as a measure of the heaviness of a distribution tail \cite{Balanda1988}, a value of $\mathcal{K}$ higher than 9 (which corresponds to a negative exponential) shows how much the tail of the considered distribution positively deviates from a negative exponential. In Fig. \ref{fig:kurt_RW_ratio}(a) we illustrated, in blue, the kurtosis of the total intensity PDFs obtained from simulations for different values of optical injection and fixed pump current $\mu=6$. Further, we displayed, in red, the ratio of extreme events occurring during each simulation according to a specific threshold of intensity. This threshold is defined in literature (see e.g. \cite{Onorato2013} and references therein) and corresponds to the mean of the total intensity plus eight times its standard deviation. Both indicators support the claim of extreme events in the system. Further, we can clearly observe that the biggest deviations from a negative exponential behavior of the data happen for low values of injection and correspond to a higher percentage of extreme events. The same is done in Fig. \ref{fig:kurt_RW_ratio}(b) for fixed injection amplitude ($E_I=0.5$) and variable pump current $\mu$ (starting close to threshold), showing that the pump current does not appear to play a significant role in extreme event occurrrence.\\ 

\begin{figure}[t!]
\centering{\subfigure[]
{\includegraphics[width=.9\columnwidth]{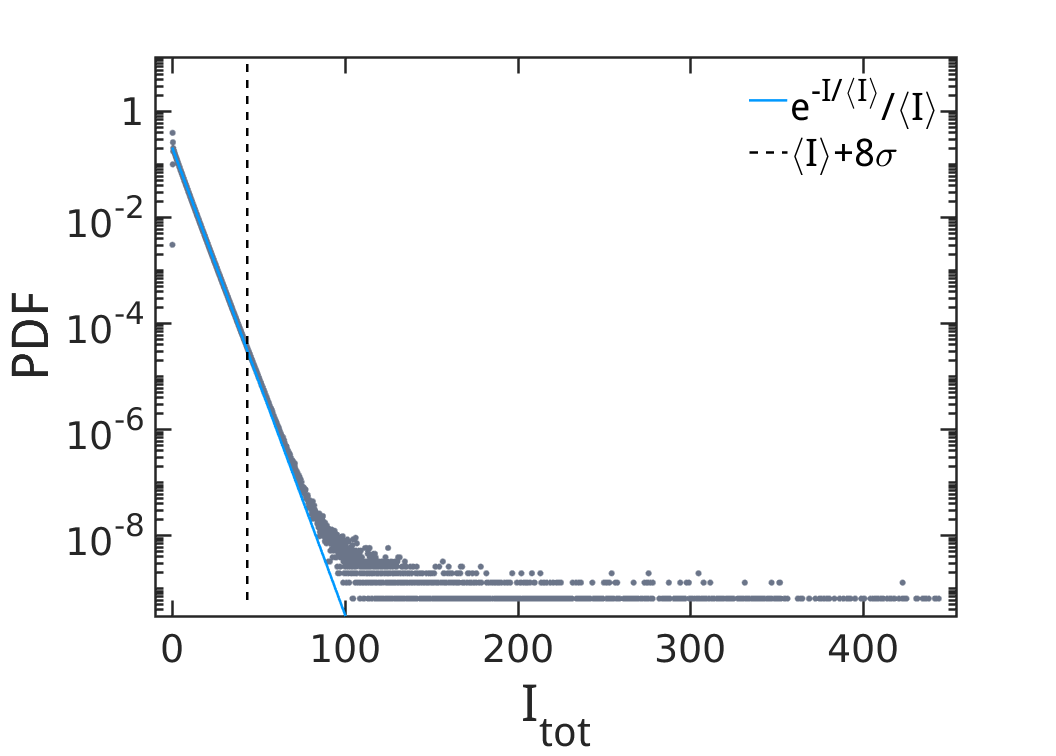}}
\subfigure[]
{\includegraphics[width=.9\columnwidth]{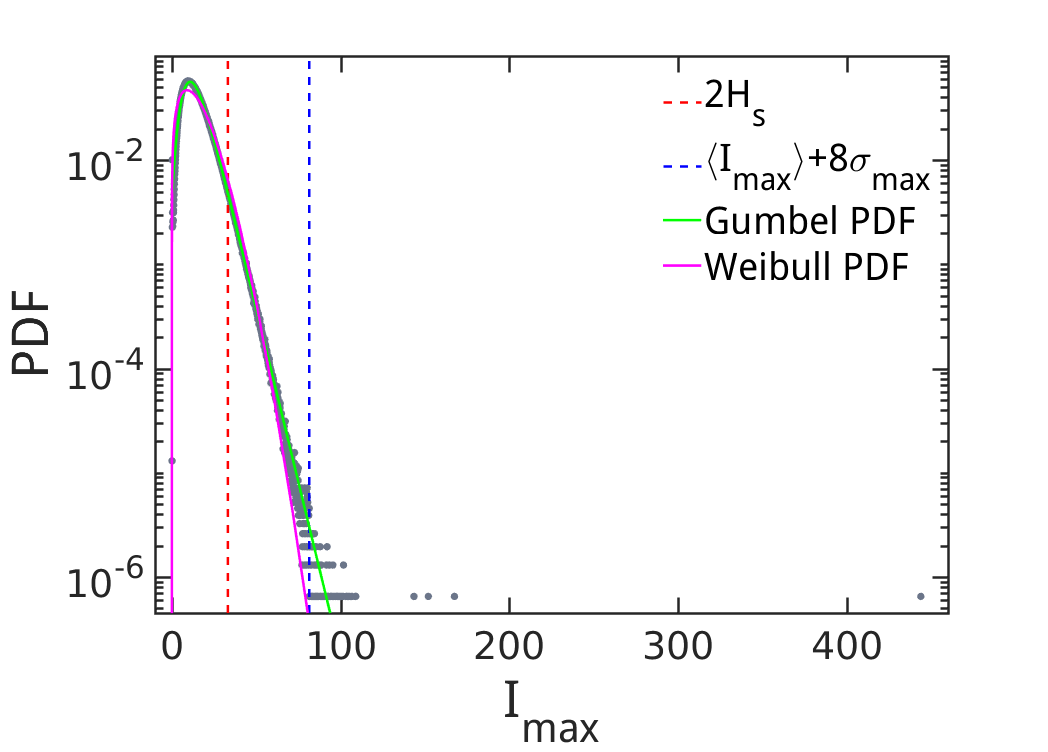}}}
\caption{Statistical analysis for a 250 ns-long simulation at $\mu=6$ and $E_I=0.1$. (a) PDF of the total intensity, where we indicated, in light-blue solid line, the negative exponential behavior which would be followed by the data in case of Gaussian statistics. (b) PDF of the spatiotemporal maxima occurring during the simulation. The magenta and green curves represent the Weibull and Gumbel PDFs, from extreme value theory. The vertical black, red, and blue dashed lines indicate the extreme event thresholds discussed in the text. Other parameters are $\alpha=4$, $\theta=-2$, $\sigma=400$ and $d=0.01$.}
\label{fig:PDFs_EI0_1}
\end{figure}

Motivated by these results, we focused on a longer simulation at $E_I=0.1$, $\mu=6$ and we reported in Fig. \ref{fig:PDFs_EI0_1}(a) the total intensity PDF for a 250 ns-long simulation ($\mathcal{K}$=9.92). We can observe the clear presence of a heavy tail in the data. Furthermore, we indicate with the vertical black dashed line the threshold for extreme events, defined as the average of the field intensity values plus 8 times their standard deviation.\\
\begin{figure}[t!]
\centering{\subfigure[]
{\includegraphics[width=\columnwidth]{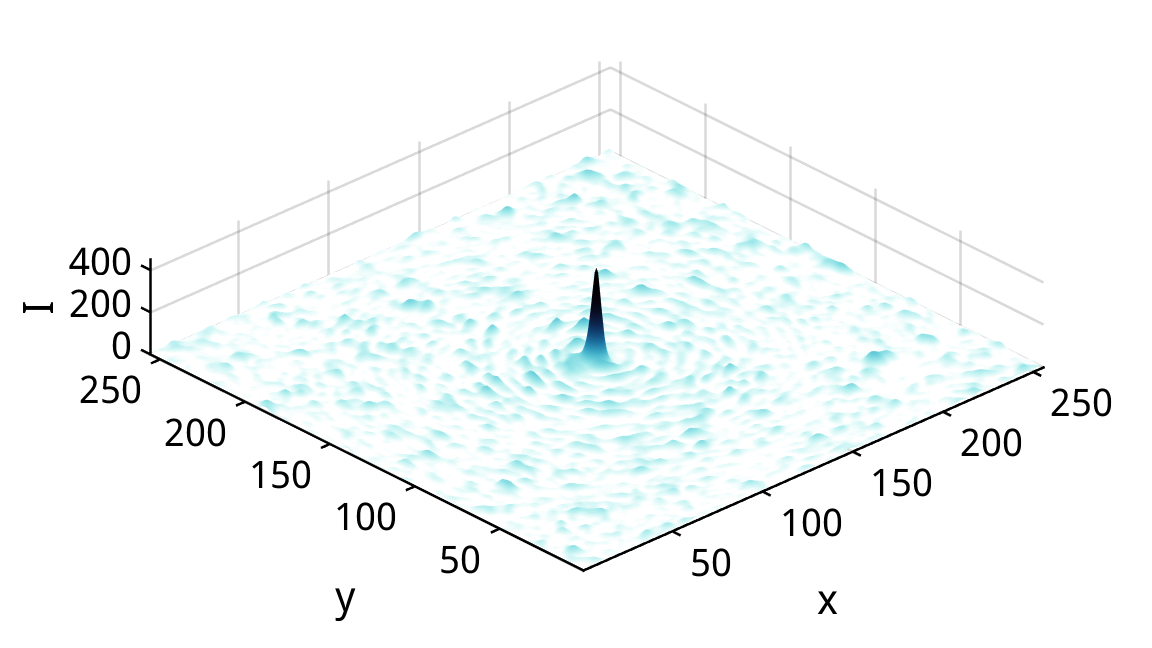}}\\
\centerline{
\subfigure[]
{\includegraphics[width=.55\columnwidth]{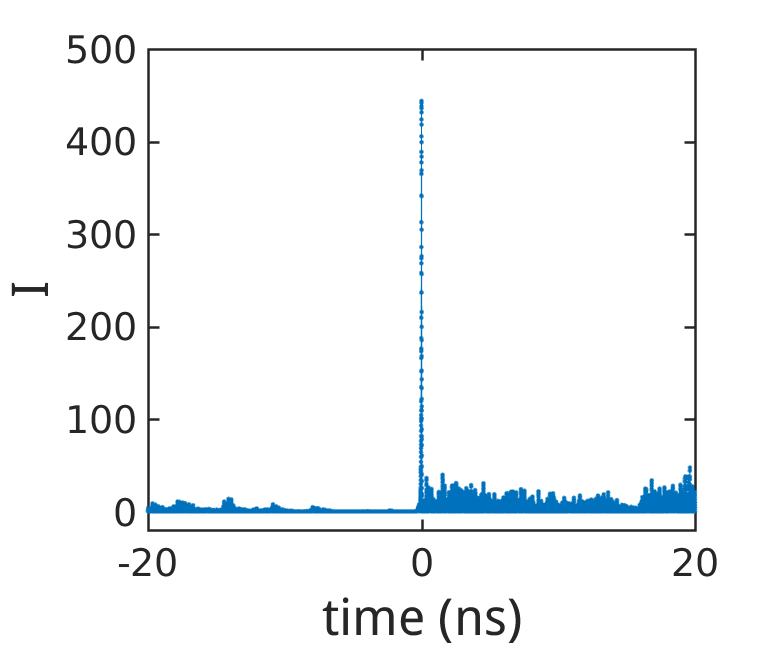}}
\subfigure[]
{\includegraphics[width=.55\columnwidth]{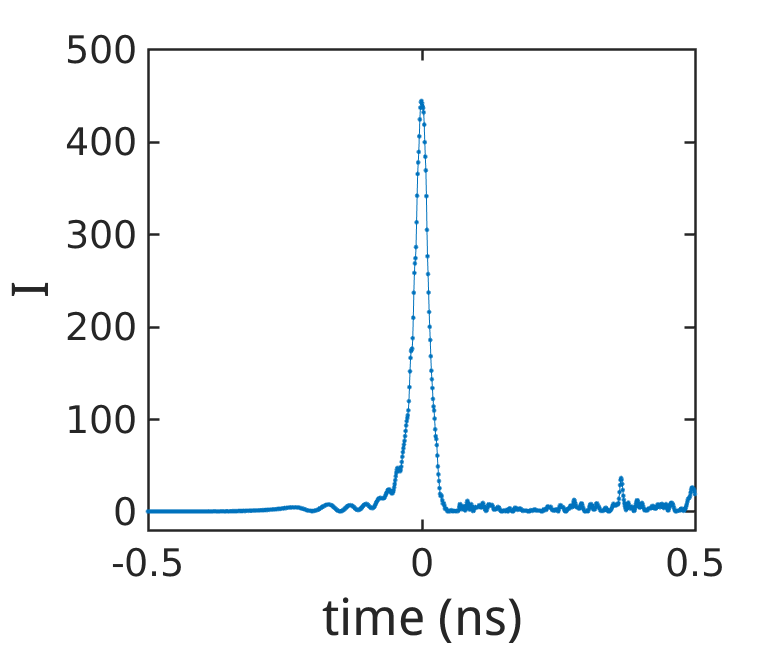}}
}}
\caption{(a) 3D intensity profile of an extreme event in the spatial transverse plane for the 250 ns-long simulation at $\mu=6$ and $E_I=0.1$ and its temporal profile (b,c). The temporal axis in (b,c) and the space grid in (a) have been centered on the extreme event maximum.}
\label{fig:EE_EI0_1_3D}
\end{figure}

In Fig. \ref{fig:PDFs_EI0_1}(b) we also illustrated the PDF of the spatiotemporal maxima occurring during the simulation. A spatiotemporal maximum is defined as a local spatial maximum in the transverse plane recorded in its temporal peak value \cite{Rimoldi2017}: a statistics performed on data sets of spatiotemporal maxima allows to identify any extreme event with a single data element, at difference with the previous analysis where multiple data points correspond to a single event. With the vertical red dashed line we reported another threshold for extreme events given by two times the significant wave height ($H_s$), that is the mean of the highest third of the maxima data. Further, we reported the more restrictive threshold, already introduced in \cite{Rimoldi2017}, given by the mean of the maxima data plus eight times its standard deviation. By means of these two thresholds, this alternative statistical study further supports the claim of extreme events in the system. The magenta and green curves represent the Weibull ($a x^{k-1}\exp(-x^k)$ with $x=I_{max}/\lambda$ and $a=k/\lambda$) and Gumbel ($\exp\{-[z+\exp(-z)]\}/\beta$ with $z=(I_{max}-\langle I_{max}\rangle)/\beta+\gamma$ and $\gamma$ Euler's constant) PDFs as families of the generalized extreme value distribution. The data is found to follow relatively well a Gumbel distribution but displays a clear deviation for very high values of intensity. 
In general, in comparison with the system studied in \cite{Rimoldi2017} (broad-area semiconductor laser with intracavity saturable absorber and no external injection), we noticed the total intensity statistics to be more sensitive than the spatiotemporal maxima statistics to the presence of extreme events.\\

\begin{figure}[t!]
\centerline{
\subfigure[]
{\includegraphics[width=.55\columnwidth]{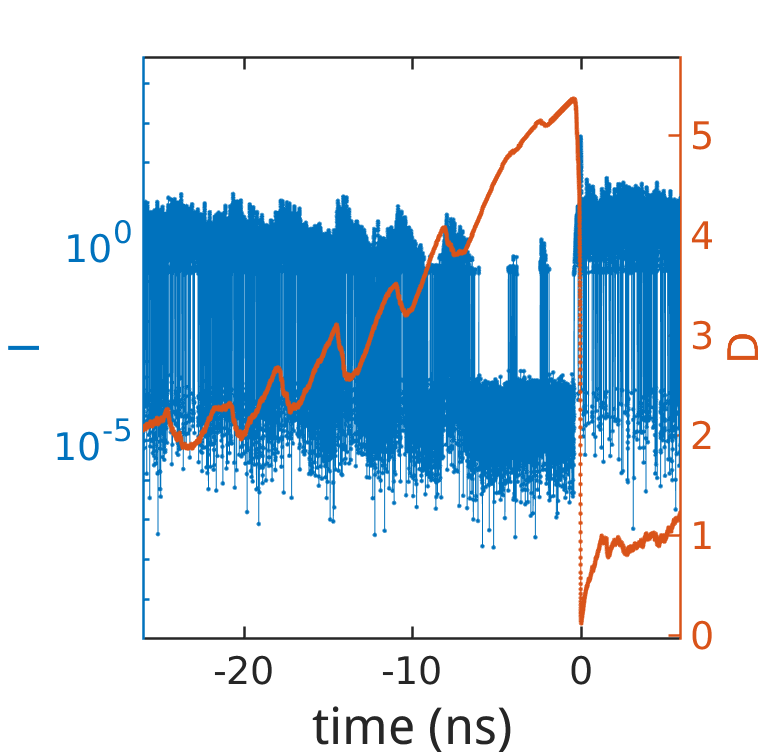}}
\subfigure[]
{\includegraphics[width=.55\columnwidth]{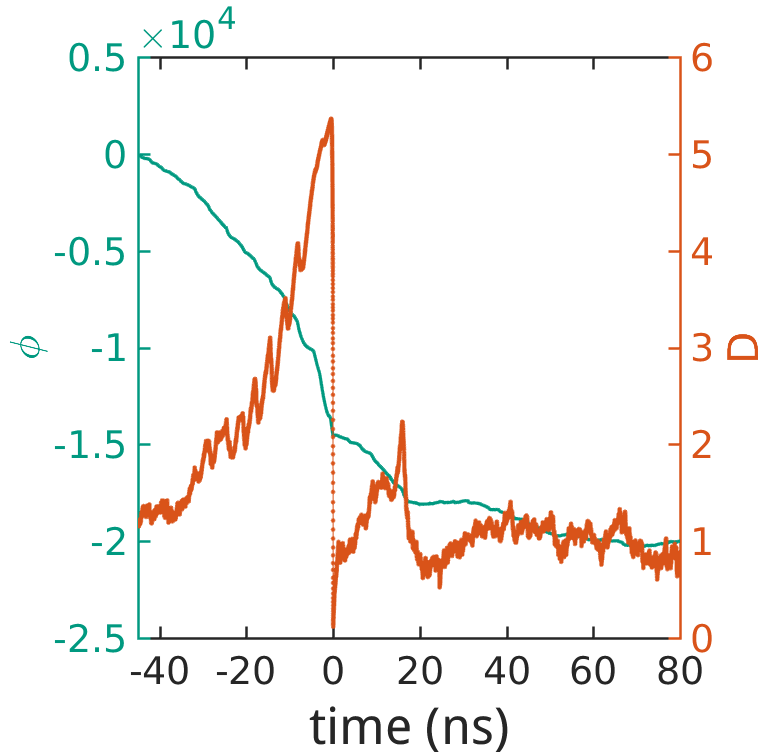}}
}
\caption{Temporal trace in a single grid point for (a) laser intensity $I$ (in logarithmic scale) and population $D$ and (b) for phase $\phi$ and population in presence of the extreme event illustrated already in Fig. \ref{fig:EE_EI0_1_3D}.} 
\label{fig:IDphi}
\end{figure}
In Fig. \ref{fig:EE_EI0_1_3D} we reported (a) the 3D profile of an extreme event occurring during the simulation at $E_I=0.1$, $\mu=6$ as well as (b,c) the (zoomed) temporal profile in the spatial grid point where the extreme event takes place (re-centered at the occurrence time). Given the periodic boundary conditions, we rearranged the spatial grid in (a) in order to visualize the extreme event at the center of the transverse plane. We can notice already in Fig. \ref{fig:EE_EI0_1_3D}(b) that prior to an extreme event the laser intensity exhibits a couple of regular pulsations onto an almost zero background. In Fig. \ref{fig:IDphi} we illustrated temporal profiles of (a) the same maximum intensity point in logarithmic scale together with that of the population variable $D$, and (b) the population at the location of the same maximum intensity point along with the (unwrapped) phase of the slowly varying envelope of the electric field.\\ 
The entire process of extreme event occurrence can be observed in several stages: (i) 20 ns before the occurrence of an extreme event the population variable starts increasing and pulsating, (ii) this progressively leads to lower values of intensity, also through a pulsating behavior, (iii) when the population variable reaches its maximum and starts decreasing, the extreme event occurs, and (iv) $D$ rapidly drops to lower values ($\approx$ 0--1). From (b) we can also observe that, even before $D$ starts increasing, the electric field phase has started to monotonically decrease (instead of oscillating around zero which is the usual behavior in other points of the grid). This dynamical description seems to be common to all the extreme events encountered for different values of pump current $\mu$ and injection $E_I$.\\

\begin{figure}[t!]
\centering 
{\includegraphics[width=.4\textwidth]{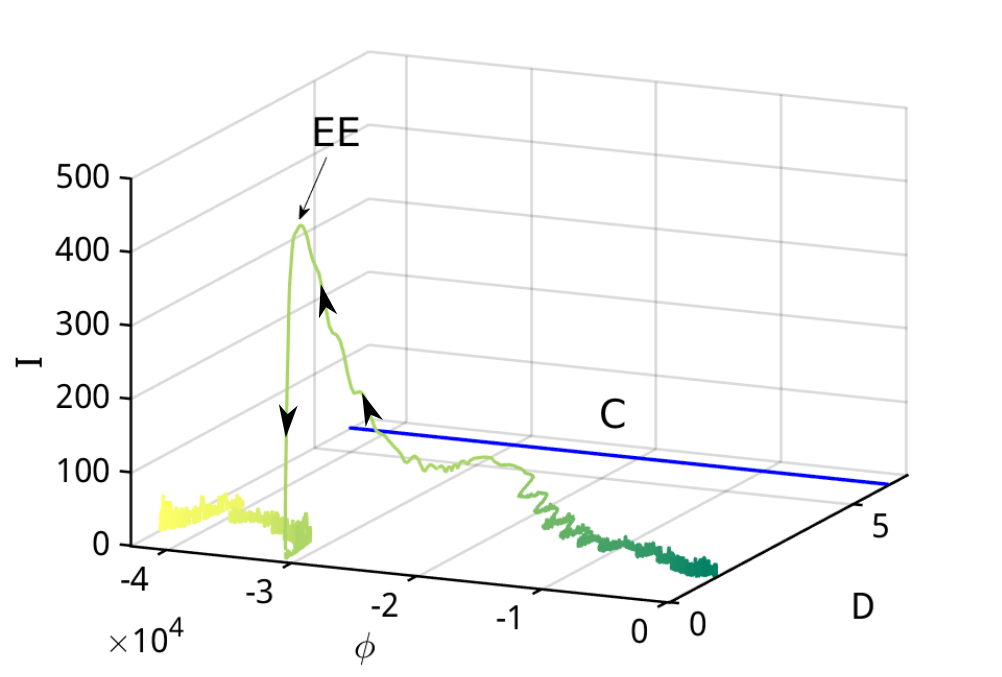}}
\caption{3D representation of the $(I,\phi,D)$ phase space for the trajectory of a single grid point following the development of an extreme event. The trajectory evolves in time from green to yellow, as also illustrated by the arrows. The blue line highlights the unstable focus $C$ of the HSS for the chosen value of injection ($E_I=0.1$). }
\label{fig:IDphi3D}
\end{figure}

In Fig. \ref{fig:IDphi3D} we illustrated the trajectory of the same point in $(I,\phi,D)$ phase space when approaching the extreme event. The blue solid line highlights the unstable focus point $C$ \cite{Rimoldi2017a} on the lower branch of the stationary curve in Fig. \ref{fig:HSS}(a) for the chosen value of the injection amplitude ($E_I=0.1$). $C$ has a specific value in phase but given the large excursion in $\phi$ even marking such value with a dot (with modulus 2$\pi$) visually results in a line along the phase axis for this scale. The trajectory evolves in time from green to yellow (in the direction of the arrows) and, as the trajectory approaches the fixed point $C$, the phase starts to decrease, the amplitude of the electric field diminishes and the population variable continues to increase. When the trajectory grows close enough to the fixed point, the repulsive nature of $C$ pushes the system to spike in intensity, which is accompanied by a rapid decrease in population. Finally, after the trajectory reaches its minimum in intensity and population, D starts to oscillate around 1. For longer times ($t>$80 ns), not illustrated in Fig. \ref{fig:IDphi3D}, the phase of the electric field starts to oscillate around a constant value, which would be zero if we take into account the 2$\pi$ modulus, as occurs in other points of the spatial grid lacking extreme events. Such a specific behavior for the phase variable allows to foresee the formation of extreme events up to 20 ns before their actual occurrence, which may potentially pave the way for their suppression through carefully thought experimental techniques \cite{Cavalcante2013}.\\

\begin{figure}[t!]
\centering 
{\includegraphics[width=.5\textwidth]{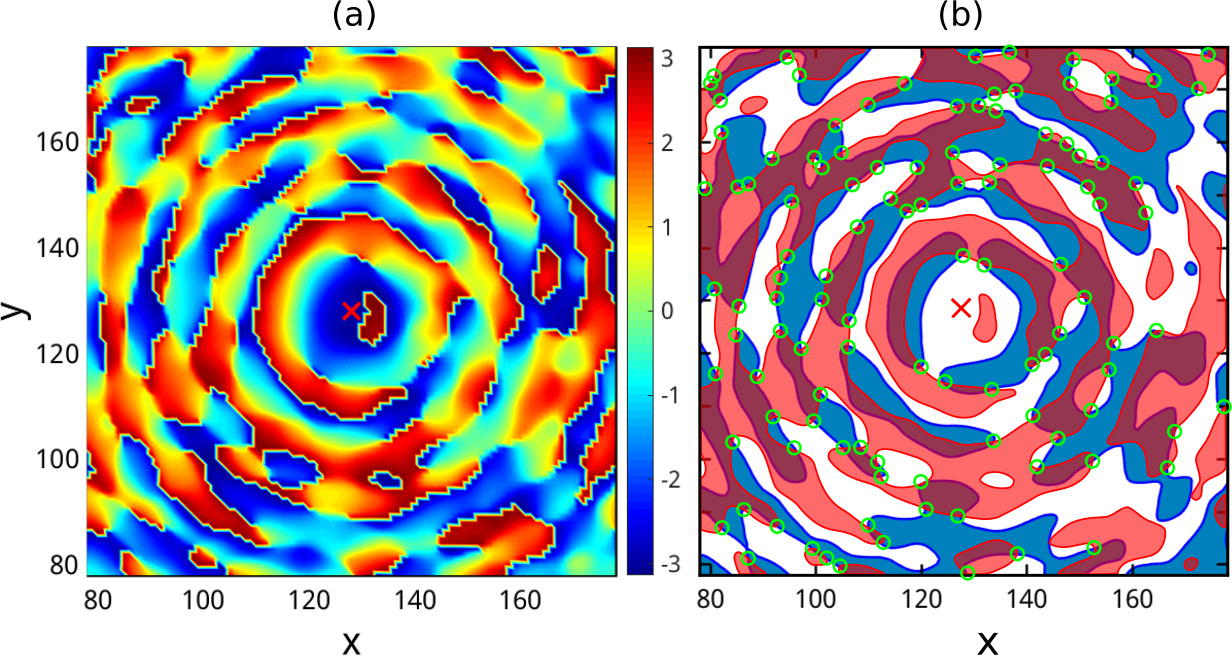}}
\caption{Phase (a) and contour plot (b) for the real (blue) and imaginary (red) part of the electric field at the time of extreme event occurrence in the spot highlighted by the red cross. In (b) the color-filled areas correspond to positive values for the respective variables, hence the border of each area identifies a zero isoline. The crossings between blue and red isolines, highlighted by green circles, identify optical vortices.}
\label{fig:phase_vortices}
\end{figure}

To further elucidate the role of the electric field phase in the dynamics, its behavior in the transverse plane at the extreme event occurrence time is shown in Fig. \ref{fig:phase_vortices}(a) where a red cross marks the location where the extreme event takes place. We can observe the presence of concentric structures around the red cross simultaneously with the occurrence of ripples in the transverse profile of the laser intensity also evident in Fig. \ref{fig:EE_EI0_1_3D}(a). It is important to observe that these ripples appear in the transverse plane long before the actual occurrence of the extreme event and are associated with the formation of vortices in the transverse plane around the extreme event future location. To better justify this point in Fig. \ref{fig:phase_vortices}(b) we illustrated a contour plot for the real (blue) and imaginary (red) parts of the electric field where color-filled areas correspond to positive values for $Re(E)$, $Im(E)$ and the border of each area corresponds to a zero isoline for the respective variable. Each crossing between blue and red isolines in Fig. \ref{fig:phase_vortices} corresponds to a zero for the laser intensity and gives rise to an optical vortex \cite{Coullet1989}, further highlighted by a green circle. Even if optical vortices (both positive and negative) occur all over the transverse plane, their specific configuration around extreme event locations seems to participate to the formation of high intensity structures.

\section{Optical gain analysis} 
\begin{figure}[t!]
\centering 
{\includegraphics[width=.45\textwidth]{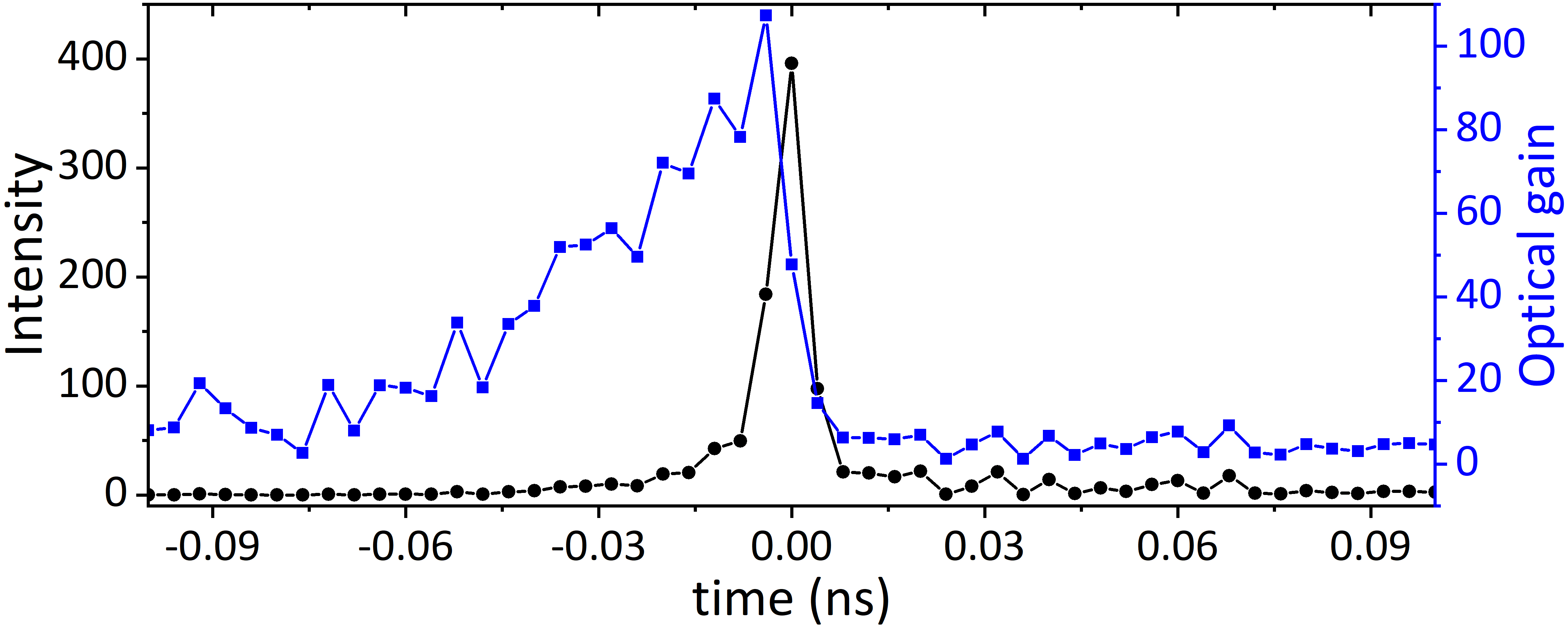}}
\caption{Intensity and optical gain values versus time at the point where the extreme event occurs (at time 0). The peak of the optical gain happens 4 ps before the extreme event. Control parameters are $\mu=8$ and $E_I=0.5$.}
\label{fig:intGtt}
\end{figure}
\begin{figure}[t!]
\centering 
{\includegraphics[width=.4\textwidth]{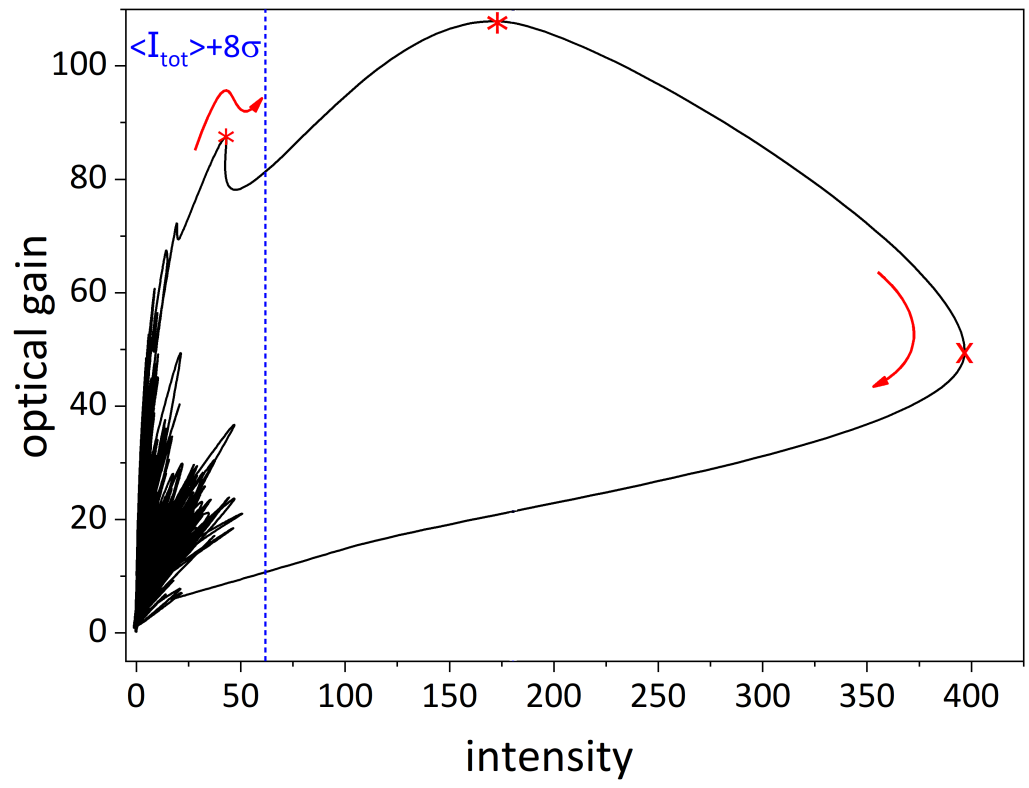}}
\caption{Trajectory of the system in the (intensity, optical gain) sub-space. When the fluctuating trend of the trajectory breaks, the relevant excursion leading to an extreme event begins. The arrows show the trajectory direction. The extreme event is marked by a red cross, while the optical gain peaks are marked by red asterisks. The vertical dashed line corresponds to the extreme event threshold as defined in Fig. \ref{fig:PDFs_EI0_1}(a). Parameters are the same as in Fig.~\ref{fig:intGtt}.} 
\label{fig:intGss}
\end{figure}

Alongside the phase and population, a closer look into the optical gain dynamics of the considered system can shed further light on the underlying mechanism of extreme event formation and its prediction. The optical gain for the model in Eq. (\ref{eq:model}) can be defined as the absolute value of $(1-i\alpha)ED$ \cite{Hachair2006}.\\ 
For a given set of control parameter values, $\mu=8$ and $E_I=0.5$, we sketched in Fig.~\ref{fig:intGtt} the temporal trace of optical gain and intensity at the spatial location of an extreme event. Here, we observe that the optical gain reaches its maximum value 4 ps before the extreme event occurrence.
To further this point, in Fig.~\ref{fig:intGss} we show the trajectory in the sub-space of intensity and optical gain. Here, the system trajectory is initially fluctuating around optical gain values of 8 (a.u.) and usually remains below 20. Then, this fluctuating behavior changes and the optical gain starts to grow beyond 70 about 22 ps before the occurrence of the extreme event, which increases the temporal range of a possible prediction window. Two clear peaks for the optical gain (marked by red asterisks in Fig.~\ref{fig:intGss}), one immediately before the event and another one after the gain starts its large excursion, can act as warnings for the upcoming extreme intensity peak. 
We can also observe that the extreme event threshold based on the values of total intensity (vertical black dashed line in Fig.~\ref{fig:intGss}) occurs after the first optical gain peak, which therefore seems more sensitive as an indicator of extreme event behavior.\\

\begin{figure}[t!]
\centering
\includegraphics[width=\columnwidth]{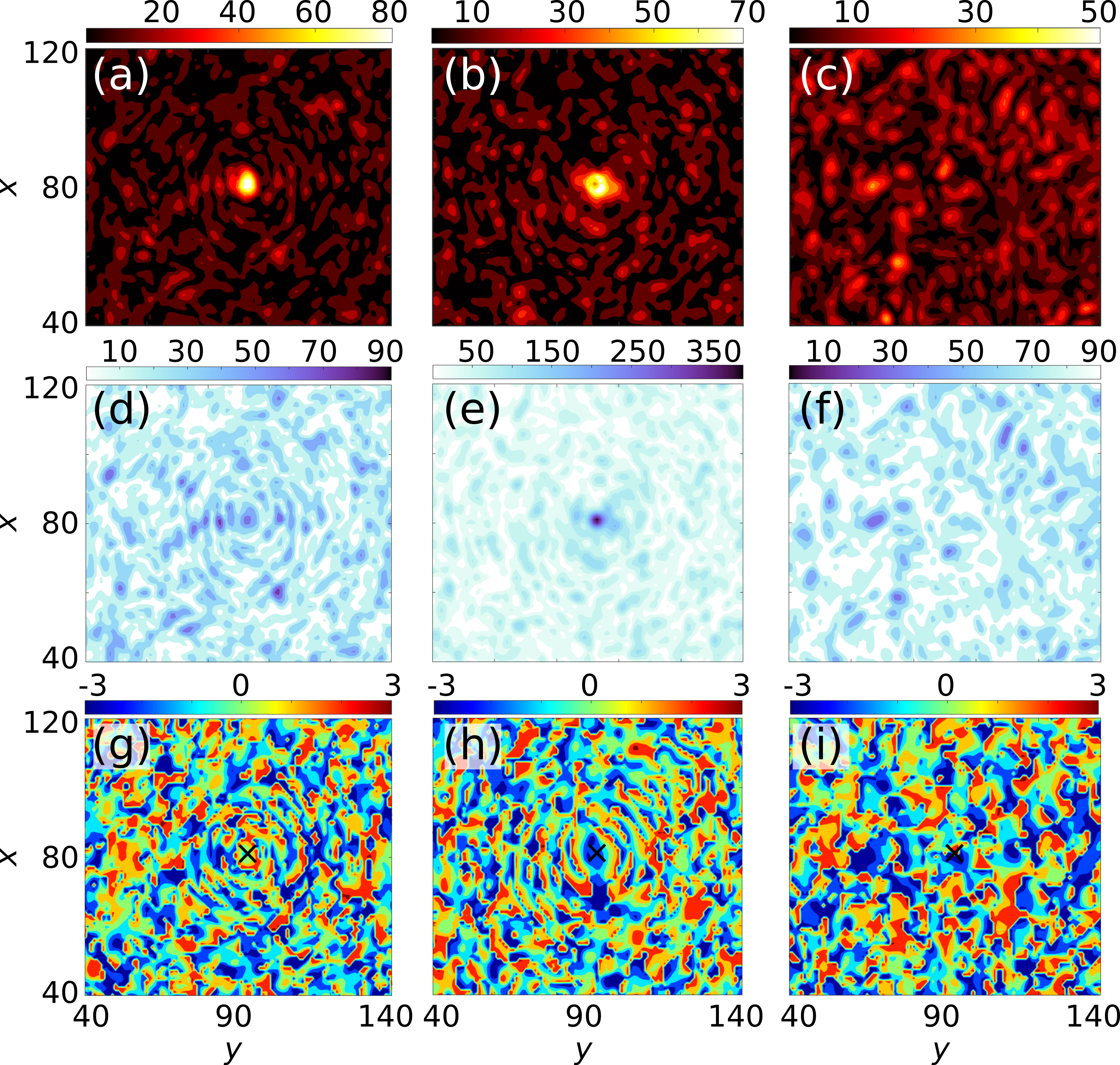}
\caption{Laser output variables in a zoom of the transverse plane at time (first column) 9.180, (second column) 9.192 and (third column) 9.204 ns. Top row: optical gain reaching its maximum in (a). Middle row: electric field intensity displaying an extreme event in (e). Bottom row: electric field phase showing the presence of vortices around the location of the event (marked with a black cross). The optical gain matrix shows a spatio-temporal maximum before the emission of the extreme event. Parameter values are the same as in Fig.~\ref{fig:intGtt}.}
\label{fig:EEplots}
\end{figure}

In order to summarize the evolution of the laser output in the proximity of an extreme event, we report in Fig.~\ref{fig:EEplots} the behavior of optical gain (first row),  electric field intensity (second row) and phase (third row) in a time window of 24 ps with an extreme event occurring at $t=12$ ps (central column). Here we can observe a peak in the optical gain occurring 12 ps before the formation of the extreme event in the electric field intensity. Note also that, while the spatiotemporal maximum in optical gain occurs 12 ps before the event, if we were to consider the beginning of the optical gain increase in the system trajectory, as illustrated in Fig.~\ref{fig:intGss}, this would allow for a larger prediction window. Vortices, visible in the phase plots in Fig.~\ref{fig:EEplots}(g,h) give rise to intensity ripples in (d,e) and this process represents a precursor of extreme event occurrence in a time window of the order of tens of ns, as discussed in the previous Section.\\
\section{Extreme events in presence of finite circular pump}
\begin{figure}[t!]
\centerline{\subfigure[]{\includegraphics[width=.45\columnwidth]{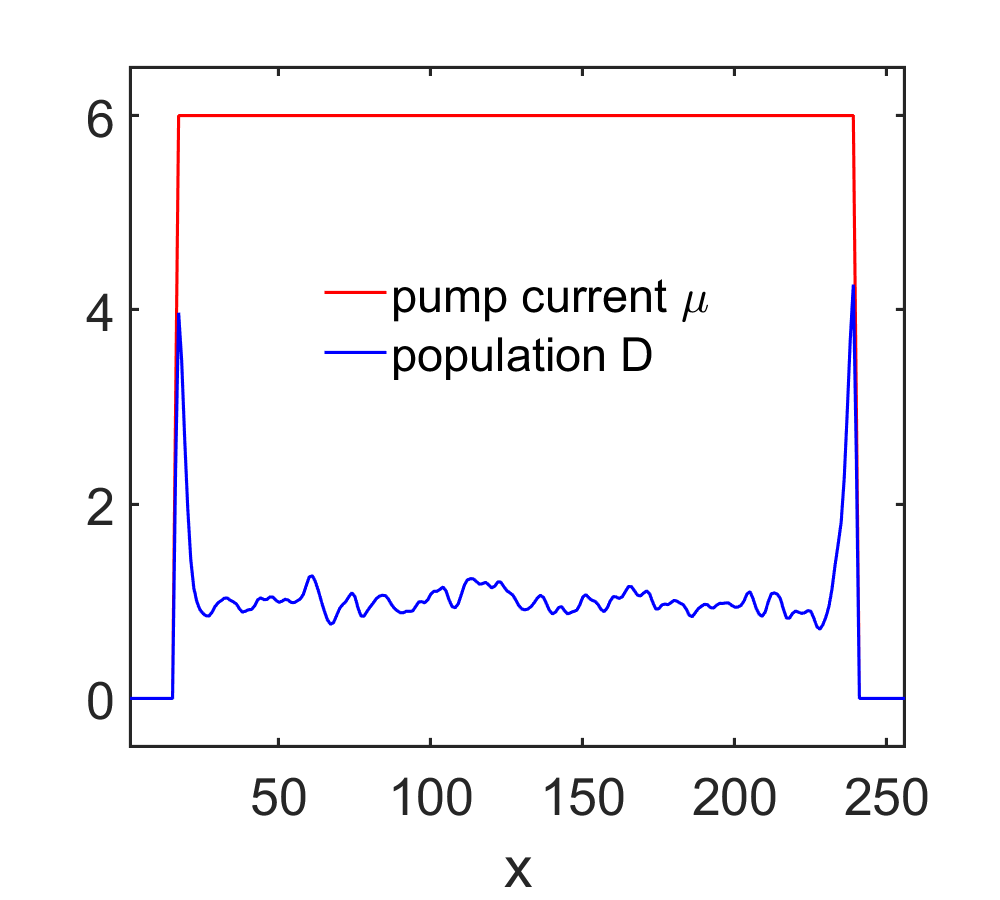}}
\subfigure[]
{\includegraphics[width=.6\columnwidth]{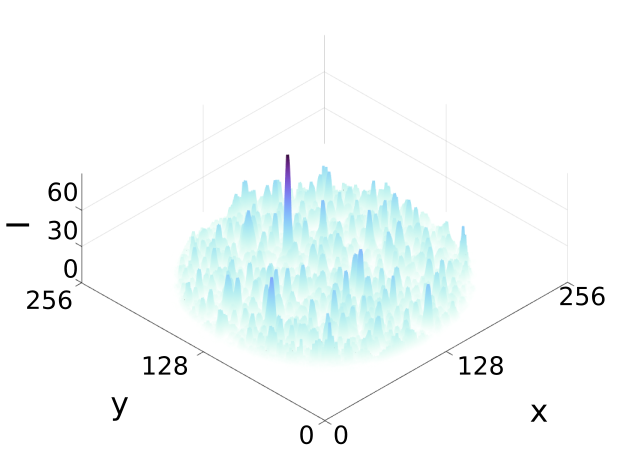}}
}
\caption{(a) Profile of the finite pump current of Eq. (\ref{eq:finite_pump}) and population at the pump diameter on the x-axis. (b) intensity snapshot at the time of extreme event formation. Parameter values are $\mu=6$ and $E_I=0.5$.}
\label{fig:Fprofiles}
\end{figure}

In all simulations we used periodic boundary conditions, which is required by the Fast Fourier Transform algorithm to solve the transverse Laplacian for the diffraction and diffusion terms. This assumption, which physically corresponds to an infinitely broad area for light emission in the transverse section of the laser, may give rise to unrealistic behaviors in simulations. Therefore, in order to confirm the results reported here, we replaced the flat pump current profile with the following function of the transverse plane coordinates, which simulates a finite circular pump profile with rapidly decaying tails
\begin{equation}
    \mu(r)=\mu/2(1-\tanh[\rho(r-r_0)])
\label{eq:finite_pump}
\end{equation}
where $\rho$ and $r_0$ regulate the size of the tail and flat part of the pump, respectively. In Fig.~\ref{fig:Fprofiles}(a) we show the form of this circular flat-top pump profile along the x-axis on a $256\times256$ grid, together with that of the population ($D$) profile at the pump diameter during simulation. In Fig.~\ref{fig:Fprofiles}(b) we show instead the transverse snapshot of the intensity at the time of an extreme event occurrence.
\begin{figure}[t!]
\centerline{
\subfigure[]
{\includegraphics[width=.7\columnwidth]{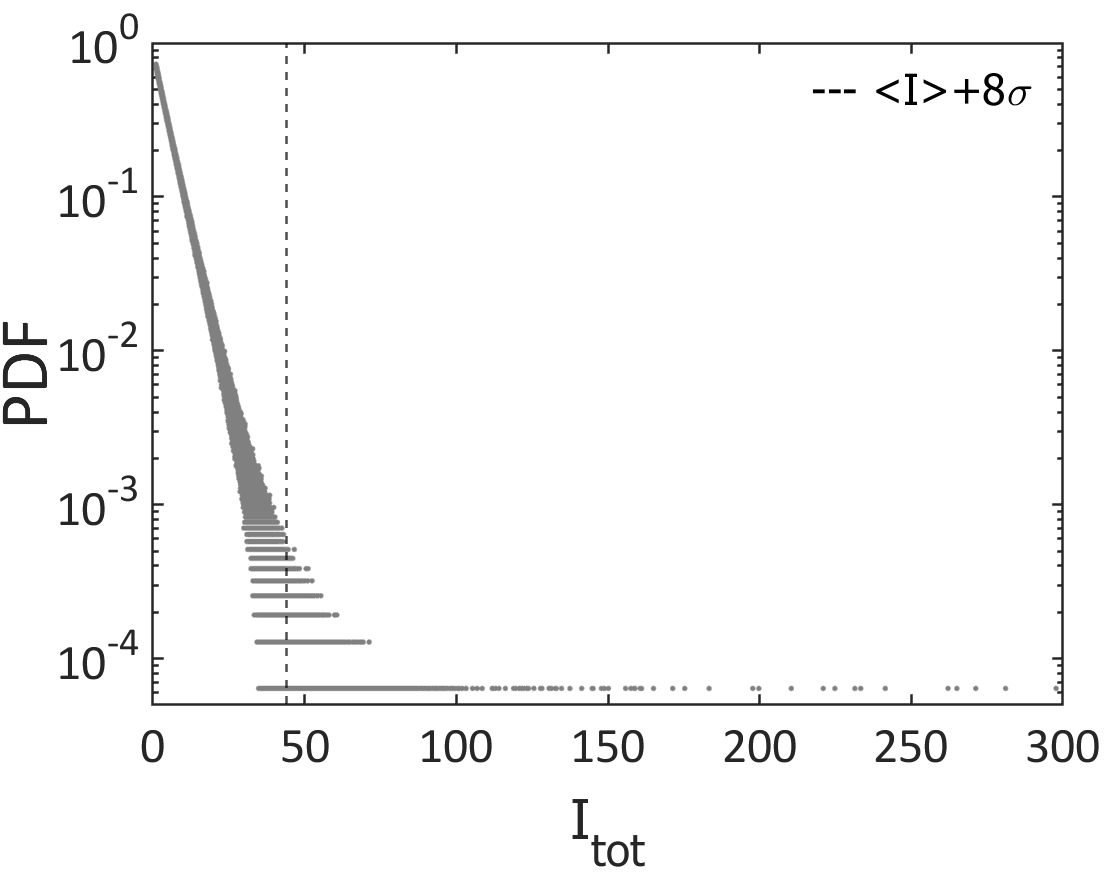}}}
\centerline{\subfigure[]
{\includegraphics[width=.7\columnwidth]{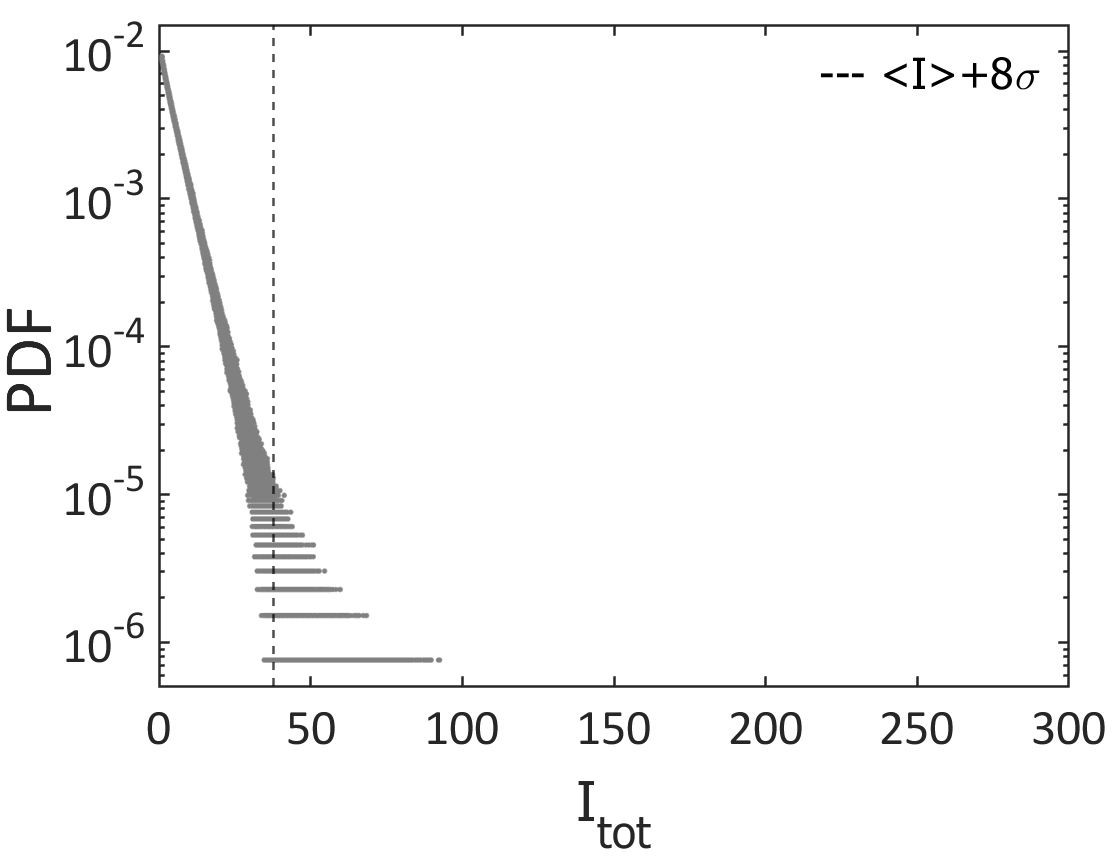}}
}
\caption{Total intensity statistics affected by finite pump, (a) PDF in case of infinite/flat pump and (b) PDF for the circular flat-top pump. We can observe extreme events still occur in presence of a finite pump profile. Parameter values are $\mu=6$ and $E_I=0.5$ in both cases.}
\label{fig:FPPDF}
\end{figure}
Performing the statistics of the total intensity data in a simulation for $\mu=6$ and $E_I=0.5$ both in the case of infinite/flat and finite circular pump profile, we obtain the results illustrated in Fig. ~\ref{fig:FPPDF}, where in the second case (b) the data outside the circular area of the pump was excluded. Here, we can observe that, while the presence of finite pump affects the shape of the statistics, it does not inhibit the presence of extreme events. Note also that while both simulations have been run for a total of 25 ns in order to compare these two PDFs beyond a qualitative level, the simulation in the finite pump case should be extended so that the number of data points considered in the statistics is the same in both cases. However, as highlighted by the figures, the deviation of the statistics from the defined threshold is still apparent, which makes all the arguments presented in this paper robust.
\section{Discussion}
In contrast with the turbulent regime in VCSELs with saturable absorber \cite{Rimoldi2017} (where extreme events were also observed), the kind of turbulence in the current system presents different features, namely the active contribution of vortices in the formation of extreme events. This makes the system dynamically closer to similar setups with smaller spatial dimensionality, where phase chirality is responsible both for the formation of phase solitons \cite{Gustave2017} and extreme events \cite{Rimoldi2017a}. Further, we may also argue that, while in the case of Ref. \cite{Rimoldi2017} spatial effects have an important role in the overall system dynamics, which translates for example in the occurrence of soliton spatial drift instability \cite{Prati2010}, in the present case, our turbulent regime seems to develop more evidently in the temporal direction. This feature locks spatiotemporal structures in a specific spatial point, at difference with \cite{Rimoldi2017}, where structures would constantly move around in the transverse plane, and allows for a meaningful temporal profile in a single spatial point of the cavity as well as clearer identification of possible predictors for extreme events.\\
While the dynamical description illustrated in Fig. \ref{fig:IDphi3D} is reminiscent of the results in \cite{Rimoldi2017a}, obtained by the authors for a semiconductor ring laser with optical injection and one spatial dimension along propagation, the main difference with the present case  consists in the specific role played by the electric field phase \cite{Barland2017}. In particular, in \cite{Rimoldi2017a} high-peak events would rise from the interplay between positive and negative chiral charges, that is $\pm 2\pi$ phase rotations, which developed in the cavity due to the simultaneous presence of three fixed points: an unstable focus in the lower branch (point $C$ in Fig. \ref{fig:IDphi3D}), a saddle on the negative slope branch, and an unstable node on the upper branch of HSS.
In the present system, where the only fixed point is on the lower branch of the HSS, the electric field phase in the location of the extreme event continuously decreases in time, giving rise to a very large number of negative charges, while optical vortices form around the event location contributing to the emergence of intensity ripples before its occurrence.\\
Optical vortices in the transverse plane have been identified as a mechanism for the formation of extreme events in \cite{Gibson2016} in a regime of vortex turbulence. In the present case, the dynamical description is more complex due to the slow population timescale and the semiconductor nature of the system. In particular, while in \cite{Gibson2016} the number of optical vortices is limited, in the present case we already noticed that vortices occur at all times and all over the transverse plane. Nevertheless, their specific configuration appears to still play a role in the formation of extreme events and is directly related to the formation of ripples in the laser intensity. Further similarities with the vortex turbulence of \cite{Gibson2016} remain to be investigated through a more in-depth regime characterization and will be addressed in future work.

\section{Conclusions}
We have observed the emergence of extreme events in a broad-area semiconductor laser with optical injection and justified their extreme nature through statistical analysis. Extreme events seem more frequent for high pump values and low values of optical injection. A detailed dynamical study on specific extreme events reveals a similar formation mechanism that involves a monotone temporal decrease in the electric field phase, a pulsating increase of the population variable and a maximum in the optical gain preceding the event, together with the emergence of optical vortices in the spatial transverse plane around the spot where the extreme event would take place. 
In particular, the process enhancing extreme events starts to be visible $\gtrsim$ 20 ns before the actual occurrence of the event. 
While a more systematic study will be necessary to further confirm these findings, the results obtained in this work allow to suggest a potential predictor for extreme events in the system.

\bibliography{LISrogue}

\end{document}